\documentclass[aps, prl, amsmath, amssymb, preprint, 
showpacs, superscriptaddress, nofootinbib]{revtex4-1}

\usepackage{graphicx}
\usepackage{textcomp}
\usepackage[utf8]{inputenc}
\usepackage{hyperref}
\usepackage{import}
\usepackage{pgf}

\hypersetup{
    colorlinks=true,       
    linkcolor=blue,          
    citecolor=blue,        
    filecolor=blue,      
    urlcolor=blue           
}

\graphicspath{{./figures/}}

\newcommand{\ket}[1]{\vert#1\rangle}

\begin{document}

\title{Analysis of a photon number resolving detector based on fluorescence readout of
an ion Coulomb crystal quantum memory inside an optical cavity}

\author{Christoph~Clausen}
\email{christoph.clausen@unige.ch}
\author{Nicolas Sangouard}
\affiliation{Group of Applied Physics, University of Geneva, CH-1211
Geneva 4, Switzerland}
\author{Michael Drewsen}
\affiliation{QUANTOP, Danish National Research Foundation Center for Quantum
Optics, Department of Physics and Astronomy, Aarhus University, DK-8000
Aarhus C., Denmark.
}
\date{\today}

\begin{abstract}
The ability to detect single photons with high efficiency is a crucial
requirement for various quantum information applications. By combining the
storage process of a quantum memory for photons with fluorescence-based quantum
state measurement, it is in principle possible to achieve high efficiency photon
counting in large ensembles of atoms. The large number of atoms can, however,
pose significant problems in terms of noise stemming from imperfect initial
state preparation and off-resonant fluorescence. We identify and analyse a
concrete implementation of a photon
number resolving detector based on an ion Coulomb crystal inside a moderately
high-finesse optical cavity. The cavity enhancement leads to an effective
optical depth of 15 for a finesse of 3000 with only about 1500 ions interacting
with the light field. We show that these values allow for
essentially noiseless detection with an efficiency larger than 93\%. Moderate
experimental parameters allow for repetition rates of about 3~kHz, limited by
the time needed for fluorescence collection and re-cooling of the ions between trials.
Our analysis may lead to the first implementation of a photon number resolving
detector in atomic ensembles.
\end{abstract}


\maketitle

\section{Introduction}
Photons have repeatedly been proved to be excellent carriers of quantum
information~\cite{Gisin2007}. As such they play important roles in experiments
that investigate the fundamental aspects of quantum mechanics, as well as in
emerging quantum technologies. The final step in many of these scenarios is the
detection of photons, making the detection efficiency a central parameter.
Additionally, the number of photons in the experiments increases steadily, and
as of today entangled states of as many as eight photons have been
created~\cite{Yao2012}. Increasing the photon number even further will be
extremely difficult without high-efficiency detectors.
At the same time, some of the most fundamental experiments with not more than
two photons have equally strong requirements.
Loophole-free tests of Bell's inequalities, for example, can ascertain the
non-local character of quantum mechanics, provided that the overall detection
efficiency is greater than $82.8\%$\footnote{It is possible
to relax the requirement on the detection efficiency down to $\eta = 2/3$ by
using non-maximally entangled states~\cite{Eberhard1993}. However, this requires
extreme signal-to-noise ratios.}~\cite{Eberhard1993}.
Still higher requirements are set by linear optics quantum computing based on
realistic single-photon sources, where scalable entanglement-generating gates
can only be achieved for a detection efficiency greater than
$90\%$~\cite{Jennewein2011}. Additionally these gates need detectors that can
distinguish a single-photon event from events with zero or multiple photons,
i.e. a basic form of photon number resolution. The ability to distinguish
different numbers of photons is an asset in many other situations.
For example, it simplifies the implementation of device independent quantum key
distribution, where the security of the key does not depend on the devices used
for its generation~\cite{Curty2011}. It also opens up new opportunities in
fundamental physics, such as the exploration of entanglement between microscopic
and macroscopic objects~\cite{Raeisi2011}.

The high demands set by quantum optics applications have in recent years
led to significant developments in single-photon detection technologies.
State-of-the-art silicon-based avalanche photo diodes have peak efficiencies of around 70\%
and low dark count rates, but currently the ability to distinguish
photon numbers remains limited~\cite{Thomas2010}.
Detectors that reach efficiencies above 90\% while at the same time maintaining a low dark count rate are still scarce~\cite{Hadfield2009}.
Only recently, close to unit efficiency and negligible dark counts have been
demonstrated with transition edge sensors~\cite{Lita2008}. While these devices
also show photon number resolution for small photon numbers, their low operating
temperature of 100~mK requires sophisticated cooling technology.

\begin{figure}
  \centering
  \includegraphics{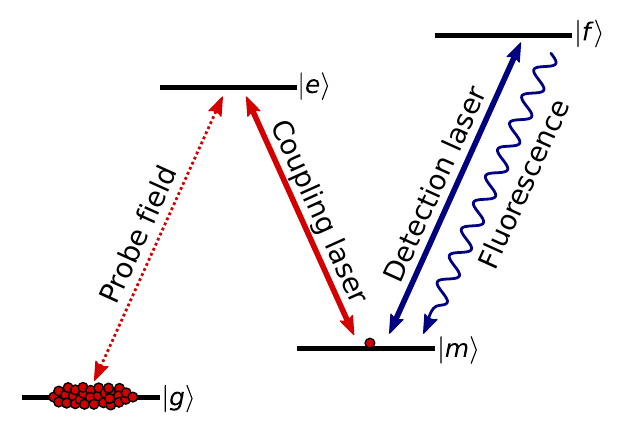}
  \caption{Principle of the atom-based photon detector as originally
  proposed~\cite{JamesKwiat2002,Imamoglu2002}. First, an ensemble of identical
  atoms is prepared in their ground state $\ket{g}$. Next, photons in the probe
  field are converted into collective excitations in the metastable state
  $\ket{m}$ with the help of a coupling laser. Finally, the number of collective
  excitations is probed by collecting fluorescence on the closed transition
  $\ket{m} \leftrightarrow \ket{f}$}
  \label{fig:original_scheme}
\end{figure}

We present a feasibility study of a high efficiency photon number resolving
detector based on a Coulomb crystal of $^{40}$Ca$^+$ ions placed inside an
optical cavity. Our scheme is a cavity-based implementation of previous proposals suggesting the use
of an ensemble of atoms to convert a single photon into many fluorescence
photons, which are then readily detected~\cite{Imamoglu2002,JamesKwiat2002}. The
fact that our cavity-based scheme only requires a moderate number of ions removes
a series of problems which would have strongly limited the usefulness of
possible implementations without a cavity. At the same time high efficiency can still
be reached with a relatively small number of
ions and a moderate finesse cavity, making faithful photon counting feasible.

The gist of the original proposals~\cite{Imamoglu2002,JamesKwiat2002} is
illustrated by means of the energy level diagram in
Fig.~\ref{fig:original_scheme}. They suggest to combine two key technologies.
First, photons in a probe pulse are coherently converted into collective
excitations in an atomic ensemble using light storage based on
electromagnetically induced transparency~(EIT)~\cite{Fleischhauer2000}. Then the number of
collective excitations is probed by measuring resonance fluorescence as usually
employed in ion trap experiments~\cite{Rowe2001}. The whole procedure works as
follows. An ensemble of atoms is initially prepared in a specific ground state
$\ket{g}$. The light field to be measured is resonant with the transition $\ket{g}
\leftrightarrow \ket{e}$. It takes the role of the probe field in an EIT scheme,
and is coherently mapped onto a collective excitation in the metastable state
$\ket{m}$ by applying a strong coupling laser on the transition $\ket{m}
\leftrightarrow \ket{e}$. Finally, a detection laser couples $\ket{m}$ to a
fourth state $\ket{f}$, which spontaneously decays back to $\ket{m}$ only. The
scheme inherently exhibits photon number resolution since
the amount of fluorescence emitted on the transition $\ket{m} \leftrightarrow
\ket{f}$ is directly proportional to the number of photons in the probe field.

The conversion of the photons in the probe field to collective excitations in
the atomic ensemble can in principle be made arbitrarily efficient by increasing
the number of atoms in the ensemble. For a cold gas the required number of atoms
is on the order of $N=10^6$. Such a large number of atoms leads to a series of
technical problems. The first problem arises during the initialization of the
atoms. Since any atom in $\ket{m}$ will contribute to the fluorescence at the
detection stage, this state has to be emptied completely, requiring optical
pumping with extremely high efficiency. For alkali atoms, considered in the
original proposals, the states $\ket{g}$ and $\ket{m}$ would typically belong to
the two hyperfine manifolds of the $S_{1/2}$ ground state, separated in energy
by a hyperfine splitting $\Delta_{\text{HFS}}$ on the order of a few
gigahertz. The state $\ket{f}$ is part of the $P_{3/2}$ manifold with a line
width $\Gamma \approx 10$~MHz. The probability of unwanted off-resonant
excitation of an atom from $\ket{g}$ during the detection stage is on the order
of $\Gamma^2/\Delta_{\text{HFS}}^2 \approx 10^{-6}$. For one million atoms this
noise is comparable to the signal from a few-photon probe field.
Finally, the collection of a sufficient amount of fluorescence can be impeded by
a premature loss of the atoms from the trap caused by heating and light-assisted
collisions~\cite{DePue1999}.

\begin{figure}
  \centering
  \includegraphics[width=\linewidth]{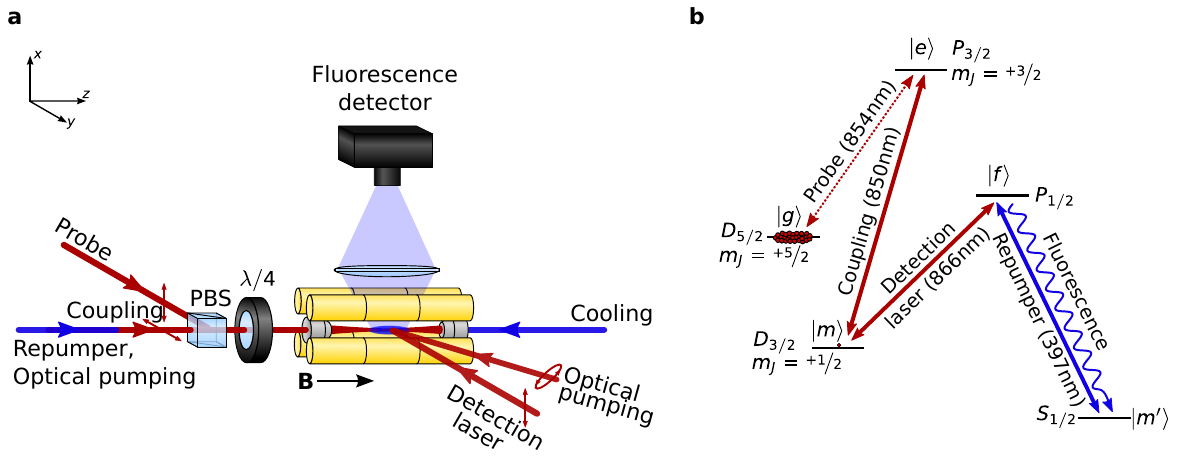}
  \caption{Implementation of the photon detector based on a 
  Coulomb crystal of ${}^{40}$Ca$^+$ ions inside an optical cavity. (a)~The
  laser beams for cooling, optical pumping, as well as the probe and coupling fields
  travel along the cavity axis in order to avoid Doppler shifts due to rf
  induced micromotion~\cite{Schiffer2000,Landa2012}. Additional laser beams are used in
  a second optical pumping step and exciting fluorescence.
  The fluorescence is collected with a large numerical aperture lens and
  directed towards a detector. (b)~Diagram that shows which
  energy levels of $^{40}$Ca$^{+}$ take the roles specified in
  Fig.~\ref{fig:original_scheme}.}
  \label{fig:ca_scheme}
\end{figure}

A concrete system which significantly reduces or avoids the problems mentioned
above is shown in Fig.~\ref{fig:ca_scheme}. It consists of an ion Coulomb
crystal~\cite{Wineland1987, Diedrich1987} with $N\approx 1500$ Ca$^+$ ions
interacting with the field of an optical cavity with a moderately high finesse
of $\mathcal{F} \approx 3000$. In the ions, the metastable states $D_{5/2}$ and
$D_{3/2}$ (lifetimes $\sim 1.15$~s) take the roles of $\ket{g}$ and $\ket{m}$,
respectively, while $P_{3/2}$ is $\ket{e}$ and $P_{1/2}$ is $\ket{f}$. Since ions in the $P_{1/2}$
$(\ket{f})$ state can spontaneously decay to $S_{1/2}$ $(\ket{m'})$, a
repumper is needed to address the $\ket{f} \leftrightarrow \ket{m'}$ transition.
The fluorescence rate on this transition is in fact more than ten times higher
than on the $\ket{f} \leftrightarrow \ket{m}$ transition, and hence most optimal
for the final fluorescence detection.

Essential ingredients of the proposed photon detector have already been
applied in recent experiments demonstrating strong collective
coupling~\cite{Herskind2009} and cavity EIT~\cite{Albert2011}.
While the experiments reported in Refs.~\onlinecite{Herskind2009,Albert2011}
were carried out between sub-states of the $D_{3/2}$ level, previously, very
efficient $(>90\%)$ coherent STIRAP population transfer between $D_{3/2}$ and
$D_{5/2}$ states had been realized~\cite{Soerensen2006}.

\section{Protocol}
In the following the individual steps of the photon detection
protocol will be discussed in detail. Every step of the protocol determines
one of the characteristics of the detector: the successful initialization
significantly reduces the probability of dark counts, the probability of
successful light storage equals the overall efficiency of the detector, and the
time required for fluorescence collection limits the repetition rate of the
detector.

\begin{figure}
  \centering
  \includegraphics[width=\linewidth]{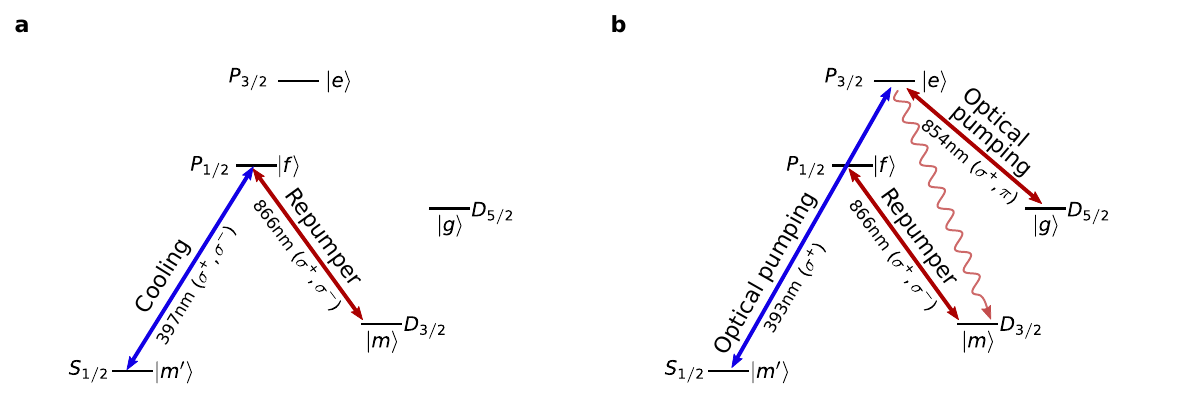}
  \caption{Energy levels and relevant transitions for the initialization of the
  photon detector in $^{40}$Ca$^{+}$. (a)~The ions are laser cooled by driving
  the $S_{1/2}\leftrightarrow P_{1/2}$ transition at 397~nm and repumping on the
  $D_{3/2}\leftrightarrow P_{1/2}$ transition at 866~nm. (b)~Optical pumping to
  the $m_J=+5/2$ Zeeman substate of the $D_{5/2}$ level is accomplished by two
  optical pumping beams on the $S_{1/2}\leftrightarrow P_{3/2}$ (393~nm) and
  $D_{5/2}\leftrightarrow P_{3/2}$ (854~nm) transitions. The repumper takes care
  of atoms that spontaneously decay from $P_{3/2}$ to $D_{3/2}$.}
  \label{fig:initialization}
\end{figure}

\subsection{Initialization}
The goal of the initialization is the preparation of a cold Coulomb crystal with
the ions in state $\ket{g}=\ket{D_{5/2}, m_J=+5/2}$.
The preparation consists of several steps similar to the preparation described
in Ref.~\onlinecite{Herskind2009}. A magnetic field of a few Gauss along the
cavity axis defines the quantization axis, and laser cooling is achieved by
applying two counter-propagating light beams along the cavity axis. The light is
resonant with the $S_{1/2}\leftrightarrow P_{1/2}$ transition at 397~nm, and the
beams are left and right-hand circularly polarized, respectively (see
Fig.~\ref{fig:initialization}). Atoms that fall into the $D_{3/2}$ state are
repumped by a laser at 866~nm applied from the side with its polarization
orthogonal to the cavity axis, equivalent to left- and right-hand circularly
polarized with respect to the quantization axis.
Once the ions are sufficiently cold, the cooling laser is turned off. Two
additional lasers pump the ions to the $m_J = +5/2$ Zeeman sublevel of the
$D_{5/2}$ state. The first of these lasers drives the $\sigma^+$-transitions
from $S_{1/2}$ to $P_{3/2}$. The second laser is resonant with the $D_{5/2}
\leftrightarrow P_{3/2}$ transition, and its propagation direction and
polarization are chosen such that $\sigma^+$ and $\pi$ transitions are addressed
simultaneously. Atoms that spontaneously decay from $P_{3/2}$ to $D_{3/2}$ are
reintroduced into the optical pumping process by the laser cooling repumper.
A typical duration of the cooling and pumping procedure is
25~µs~\cite{AlbertThesis}.

Efficient optical pumping is very important for high fidelity measurements of
the photon number in the probe field. Ions that are not in state $\ket{g}$ after
the initialization may offset the fluorescence in the final step of the detector
protocol, leading to an overestimation of the photon number.
Ref.~\onlinecite{Herskind2009} states an optical pumping efficiency to the
$D_{3/2}$ level with $m_J=+3/2$ of 97\%. Optical pumping into the $m_J=+5/2$
sub-state of the $D_{5/2}$ state is expected to have a similar efficiency. The
efficiency is limited by imperfect polarization of the pumping light, leading to
a distribution of the remaining ions over the other $D_{5/2}$ sub-states. The
spontaneous decay rate into $D_{3/2}$ of only $2\pi \times 0.18$~MHz is very
weak compared to the pumping and repumping fields, so only a tiny fraction of the
ions will end up in $S_{1/2}$ or $D_{3/2}$. The number of these ions can be estimated by
monitoring the ultraviolet fluorescence during the optical pumping, and the
result subtracted from the photon number measurement at the end of the protocol.
Alternatively, one can extend the dead time of the detector by a variable amount
and let optical pumping proceed until the moment when the monitored fluorescence
ceases. This signals that the relevant energy levels are empty, and the detector
is ready to receive the probe pulse. We note that the larger the total number of
ions, the more difficult it is to transfer \emph{all} ions to $D_{5/2}$, making
a moderate number of ions the preferred choice.

\subsection{Light storage}
In the second step of the detector protocol, the photons in the probe pulse are
converted into collective excitations in the metastable state $\ket{m}$. The
procedure is the same as the absorption of photons into a quantum memory for
light, based on an ensemble placed inside an optical cavity~\cite{Lukin2000}.
We consider a storage scheme based on EIT, where the probe pulse is resonant with the
$\sigma^{-}$-transition $\ket{D_{5/2}, m_J=+5/2} \leftrightarrow \ket{P_{3/2}, m_J=+3/2}$.
At the same time a strong coupling field is acting on the transition 
$\ket{P_{3/2}, m_J=+3/2} \leftrightarrow \ket{D_{3/2}, m_J=+1/2}$ (see also Fig.~\ref{fig:ca_scheme}).
The strength of the coupling field determines the width of the EIT window and the group
velocity of the probe pulse. One can coherently convert the quantum state of the probe pulse into a
collective excitation by reducing the group velocity to zero, that is, by adiabatically turning
off the coupling field. 

The most essential parameter of such memories is the cooperativity $C = g^2 N /
\kappa \gamma$, where $g$ is the coupling rate between a single ion and a
cavity photon, $N$ is the \emph{effective}\footnote{The total number of ions may be a
factor of 10 larger.} number of ions interacting with the mode of
the cavity~\cite{Herskind2009}, $\kappa$ is the cavity decay rate and $\gamma$
is the rate of decoherence on the transition $\ket{g} \leftrightarrow \ket{e}$
in the protocol. The cooperativity determines the maximally obtainable
photon conversion efficiency $\eta=C/(1+C)$~\cite{Gorshkov2007}. In principle, 
probe pulses of any temporal shape can be stored with optimal efficiency by adapting the
shape of the coupling field. However, the adiabatic conversion of photons
into collective excitations requires that the temporal length $T$ of the
photonic probe pulse is much larger than $1/C\gamma$~\cite{Gorshkov2007,Zangenberg2012}.
Optimal storage and retrieval with an efficiency of $\eta^2 \simeq 45\%$
has been demonstrated using EIT in Rubidium vapor~\cite{Phillips2008} without cavity.
Using a low-finesse cavity, a retrieval efficiency of 73\% was recently obtained with cold atoms~\cite{Bao2012}
using a Raman-scheme that has the same adiabaticity conditions as EIT.
   
The efficiency of the light storage determines the overall detection efficiency,
as long as saturation effects are avoided by ensuring that the number of photons
in the probe pulse is much smaller than the number of ions. The experimental
parameters obtained in Ref.~\onlinecite{Herskind2009} on the transition
$\ket{D_{3/2}, m_J=+3/2} \leftrightarrow \ket{P_{1/2}, m_J=+1/2}$ are
($g=2\pi\times 0.53$~MHz, $N\simeq 1500$, $\kappa=2\pi\times 2.15$~MHz,
$\gamma=2\pi\times 11.9$~MHz), giving a maximum cooperativity\footnote{Please note that
definition of the cooperativity in \cite{Herskind2009} differs from the one used
here by a factor of 2.} of $C\simeq 16$.
For the transitition $\ket{D_{5/2}, m_J=+5/2} \leftrightarrow \ket{P_{3/2},
m_J=+3/2}$ used in our protocol, the theoretical values for $g$ and $\gamma$ are 
slightly higher. Hence, a cooperativity of $C \simeq 15$ should be straightforward
to obtain, which gives a detection efficiency above 93\%.

The duration of the storage process is essentially given by the duration of
the probe pulse. We find that $C\gamma \approx 10^9\ \text{s}^{-1}$, allowing, in principle,
for the optimal storage of Fourier-limited pulses of duration down to about
$T\simeq 50/C\gamma =50$~ns~\cite{Gorshkov2007}, but even a more conservative
value of 1~µs is negligible compared to the duration of the entire protocol.

To avoid unwanted Doppler efffects related to rf induced micromotion (see
references in the caption to Fig.~\ref{fig:ca_scheme}), the coupling field has
to co-propagate with the probe field inside the cavity, as indicated in
Fig.~\ref{fig:ca_scheme}. In this case the geometrical constraints on the
spatial profile of the coupling field reduce the optimal storage efficiency by
an amount that depends on the radial extension of the Coulomb crystal. However,
this reduction can be rendered negligible by carefully choosing the size of the
crystal, adjusting the strength of the coupling field, or constraining the
radius by addition of a second calcium isotope~\cite{Zangenberg2012}.

\subsection{Fluorescence collection}
In the last step of the protocol, the number of ions transferred to the
$D_{3/2}$ state is measured by fluorescence collection.
Fluorescence collection is routinely applied in trapped-ion based quantum
computing, and a single-ion state-discrimination with an error probability below
$10^{-4}$ has been reported~\cite{Myerson2008}. The fluorescence is induced by
the same lasers that were used during the cooling stage
(Fig.~\ref{fig:initialization}a).
Ions in one of the the $D_{3/2}$ will emit fluorescence on the
$S_{1/2}\leftrightarrow P_{1/2}$ transition. The amount of fluorescence is
proportional to the number of ions undergoing the optical cycling. For unit
absorption efficiency and an ideal initialization of the detector, this number
is equal to the number of photons originally present in the probe pulse.
Since the detection laser is 12~nm detuned from the $D_{5/2} \leftrightarrow
P_{3/2}$ transition, the ions that remained in $D_{5/2}$ are not affected at
all. However, because the lifetime of the $D$-states is finite ($\tau_D = 1.15$~s),
ions in $D_{5/2}$ can still enter the fluorescence cycle by sponteously decaying
into $S_{1/2}$.

\begin{figure}
  	\begin{center}
  	\includegraphics{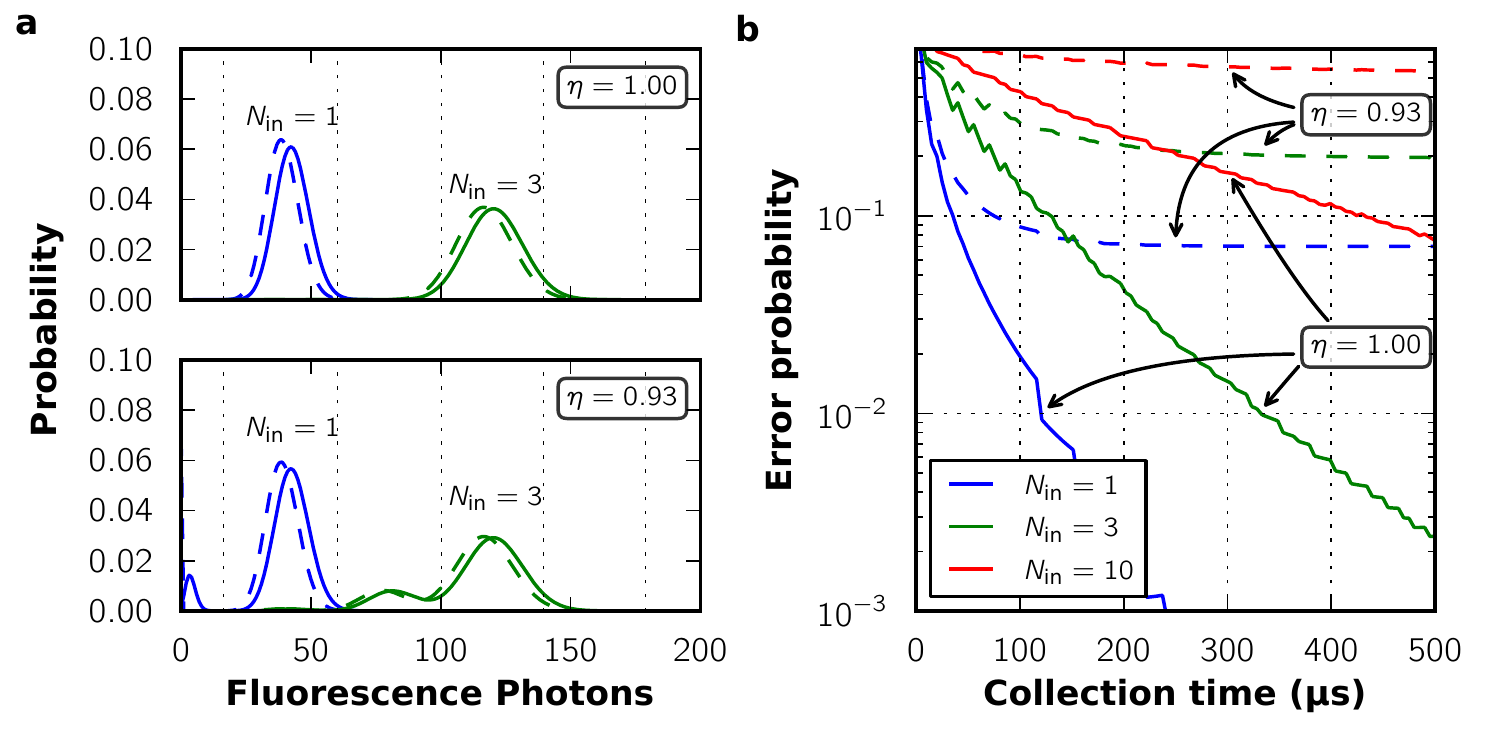}
  	\end{center}
  	\caption{Estimation of input photon number and associated error probability.
  		(a)~Fluorescence photon number distribution for $N_\text{in}=1$ and
  		$N_\text{in}=3$ input photons for a collection time of 150~µs. The upper
  		plot shows the case of ideal conversion efficiency. Dashed lines are without
  		the contribution from the spontaneous decay of ions from the $D_{5/2}$ state.
  		Vertical dashed lines indicate the thresholds for photon number estimation.
  		In the lower plot the finite conversion efficiency ($\eta=0.93$) leads to
  		the possibility to estimate a photon number lower than $N_\text{in}$. (b)~Probability of estimating a
  		photon number different from $N_\text{in}$, for $N_\text{in}=1, 3$ or 10.
  		The error probability decreases with collection time, but has a lower
  		limit for finite conversion efficiency.}
  	\label{fig:photon_number_error}
\end{figure}

We now analyse the process of fluorescence collection in a more quantitative
way. The Poissonian statistics of the detected fluorescence photons is taken
into account, and the fidelity of the photon number estimation discussed. 
Let us start by considering $\eta = 1$ and $N_\text{in}$ photons in the probe pulse. 
Assuming that all the involved optical transitions are satured, every ion spends
about 1/4 of its time in $P_{1/2}$, from where it spontaneously decays into
$S_{1/2}$ at a rate of $\gamma_{PS} = 2\pi \times 20.7$~MHz. Letting $\Theta$
denote the amount of solid angle covered by the collection system, and $\eta_D$
the overall detection efficiency at the relevant wavelength of 397~nm, the
photon detection rate per ion is given by $R = \gamma_{PS} \Theta \eta_D / 16
\pi$. A typical detector has $\eta_D = 0.4$, and a lens with 4~cm diameter
at a working distance of 7~cm can cover about $\Theta/4\pi = 2\%$ of the
full solid angle and still image the whole crystal, albeit with some distortion.
These parameters give $R=260$~kHz. The total number of photons collected
after a time $t$ will follow a Poissonian distribution with mean
\begin{equation}
	\mu_\text{in}(t) = N_\text{in}\,R\,t. 
\end{equation}
At any intermediate time $t'$, the mean number of ions having decayed
from $D_{5/2}$ is $N(1-e^{-t'/\tau_D})$, neglecting the time passed since the
initial state preparation. After a time $t$, the mean number
of collected fluorescence photons from these ions is
\begin{equation}
  	\mu_\text{decay}(t) = \int_0^t N(1-e^{-t'\tau_D}) R\, dt'
  	= N\, R\, \tau_D \left[(e^{-t/\tau_D} - 1) + t/\tau_D \right]. 
\end{equation}
The probability of detecting $N_\text{fl}$ fluorescence photons after a time
$t$ is then given by
\begin{equation}
  \label{eq:fluorescence_distribution}
  	p_{N_\text{fl}}(t) = \sum_{n=0}^{N_\text{fl}} \text{Po}[n; \mu_\text{in}(t)]
  	\cdot \text{Po}[N_\text{fl}-n; \mu_\text{decay}(t)]
  	= \text{Po}[N_\text{fl}; \mu_\text{in}(t) + \mu_\text{decay}(t)],
\end{equation}
where $\text{Po}[n; \mu]$ denotes the Poisson distribution with mean $\mu$.
So the spontaneously decaying ions shift the amount of fluorescence to slightly
higher values without changing the shape of the distribution. This is illustrated
in Fig.~\ref{fig:photon_number_error}a for $N=1500$ ions. 

The time required for fluorescence collection depends on the amount of
confidence on the estimated input photon number that one wants to obtain. We
consider a simple strategy, where the input is estimated to $N_\text{in}$
photons if the amount of fluorescence is larger than a threshold that
corresponds to the point where the Poisson distributions for $N_\text{in}$ and
$N_\text{in}-1$ input photons cross, see Fig.~\ref{fig:photon_number_error}a.
As a measure of the fidelity we will consider the probability that $N_\text{in}$
input photons lead to an estimate that different from $N_\text{in}$. This error probability
is plotted as a function of the collection time in Fig.~\ref{fig:photon_number_error}(b)
for $N_\text{in} = 1,3$ or~10 photons. The larger $N_\text{in}$, the more fluorescence
is needed to reduced the error below a certain level. For the parameters considered here
and an error probability below 10\%, one can distinguish up to 3 photons after
$t\simeq 130$~µs, and 10 photons after $t\simeq 430$~µs.

In the case where $\eta < 1$, the number of ions transferred to $\ket{m}$ follows
a binomial distribution, that is, not all photons are necessarily converted into collective
excitations. In fact, the probability that \emph{all} input photons are converted
is $\eta^{N_\text{in}}$, which sets a lower bound on the error probability of
$p_\text{err}^\text{min} = 1 - \eta^{N_\text{in}}$. We note that this lower bound
is valid for \emph{any} photon number resolving detector with non-unit efficiency.
For our parameters, the bound is obtained after $t\simeq 180$~µs for $N_\text{in}=1$, and
$t\simeq 250$~µs for $N_\text{in}=3$.

After the fluorescence detection, the ions will have to undergo a brief
period of laser cooling before reinitialization back into $D_{5/2}$. Based on
previous experiments~\cite{Herskind2009,Albert2011}, this procedure is expected
to take less than 100~µs.

To calculate the repetition rate, we add up the durations of the individual steps of
the protocol, neglecting the short duration of the light storage. Using the numbers
stated in the previous sections, we get $T_\text{total} \simeq (25 + 200 + 100)$~µs,
giving a repitition rate of about $3$~kHz.

\section{Conclusion}
In summary, we have presented a concrete implementation of an
atomic-ensemble-based photon number resolving detector based on a Coulomb
crystal of $^{40}$Ca$^+$ ions inside an optical cavity. For a currently
available system, a detection efficiency of $\eta \approx 93\%$ is already
feasible. The efficiency can be improved by increasing the cooperativity, e.g.
by applying a cavity with a higher finesse, a larger Coulomb crystal, and/or
spatially controlling the ions positions with respect to the anti-nodes of the
standing-wave light field~\cite{Linnet2012,Enderlein2012}.
A detection efficiency of better than $98\%$ is thus within reach.
Different photon numbers can be distinguished as long as the number of
photons is much lower than the number of ions. For 1500 ions, photon number
resolution can probably be maintained up to a few tens of photons. However, for
more than $\sim 10$ input photons the non-unit detection efficiency and
the Poissonian counting statistics of fluorescence will limit the achievable
fidelity. Furthermore, it should be possible to reduce the dark counts to a negligible
level, provided that the quality of the initialization can be assured.
On the downside, the detector can be considered as rather slow in terms of repetition rate.
Currently, the repetition rate would be limited to about 3~kHz, caused by the
rather long fluorescence detection and cooling steps. Improvements of the experimental setup could
probably push this to about one measurement every 50~µs. This would correspond
to the time it takes for a signal to travel through 10~km of optical
fibre, and should hence be sufficient for, e.g., the next generation of
long-distance quantum communication experiments~\cite{Sangouard2011}. The
bandwidth of the photons in the probe pulse is limited by the adiabaticity condition $T \gg 1/C\gamma$
for the light storage. For our proposed implementation using $^{40}$Ca this
translates into a minimum Fourier-limited duration of the input pulse of about 50~ns.
The improvements of the cooperativity discussed above can probably
gain a factor of 4 to 5.  However, many tasks in optical quantum information
processing also necessitate quantum memories, whose bandwidth is equally
limited. In fact, the kind of photon detector presented here is a quantum memory
without retrieval, and it is imaginable that other kind of quantum memories can
be used in a similar way, extending the range of applications for quantum
memories significantly.

\begin{acknowledgments}
We are grateful for discussions with Mikael Afzelius, Aurelien Dantan, Nicolas
Gisin, Philippe Goldner, Jean-Louis Le Gouët, Rob Thew and Hugo Zbinden. CC and NS acknowledge
financial support through the ERC Advanced Grant QORE and from the Swiss NCCR ``Quantum Science and Technology'' (QSIT). MD acknowledges financial support from the Carlsberg Foundation and the EU via the FP7 projects ``Physics of Ion Coulomb Crystals'' (PICC) and ``Circuit and
Cavity Electrodynamics'' (CCQED).
\end{acknowledgments}

\bibliographystyle{apsrev4-1}
\bibliography{fluorescencedetector}

\end{document}